\documentclass[12pt,final]{l4dc2020}

\title[NeurOpt]{NeurOpt: Neural network based optimization for building energy management and climate control}
\usepackage{times}

\usepackage{hyperref}
\usepackage[font=small]{caption}
\usepackage{algpseudocode}
\usepackage{booktabs}
\usepackage{todonotes}
\usepackage[per-mode=symbol]{siunitx}
\usepackage{tikz}
\usepackage[utf8]{inputenc}
\usepackage{mathtools}  
\usepackage{xspace}

\usepackage{changes}
\definechangesauthor[name={F S}, color=red]{fs}

\DeclareMathOperator*{\minimize}{\text{minimize}}

\newcommand{\st}{\mbox{\text{subject to}}}

\newcommand{\customspace}{\vspace{3pt}}
\author{
 \Name{Achin Jain} \Email{achinj@seas.upenn.edu}\\
 \addr University of Pennsylvania, Philadelphia 19104, PA, USA
  \AND
 \Name{Francesco Smarra} \Email{francesco.smarra@univaq.it}\\
 \addr Università degli Studi dell'Aquila, L'Aquila 67100, Italy
  \AND
 \Name{Enrico Reticcioli} \Email{enrico.reticcioli@graduate.univaq.it}\\
 \addr Università degli Studi dell'Aquila, L'Aquila 67100, Italy
  \AND
 \Name{Alessandro D'Innocenzo} \Email{alessandro.dinnocenzo@univaq.it}\\
 \addr Università degli Studi dell'Aquila, L'Aquila 67100, Italy
 \AND
 \Name{Manfred Morari} \Email{morari@seas.upenn.edu}\\
 \addr University of Pennsylvania, Philadelphia 19104, PA, USA
}

\begin{document}

\maketitle


\begin{abstract}
	Model predictive control (MPC) can provide significant energy cost savings in building operations in the form of energy-efficient control with better occupant comfort, lower peak demand charges, and risk-free participation in demand response. 
	However, the engineering effort required to obtain physics-based models of buildings is considered to be the biggest bottleneck in making MPC scalable to real buildings.
	In this paper, we propose a data-driven control algorithm based on neural networks to reduce this cost of model identification. 
	Our approach does not require building domain expertise or retrofitting of existing heating and cooling systems.
	We validate our learning and control algorithms on a two-story building with ten independently controlled zones, located in Italy. 
	We learn dynamical models of energy consumption and zone temperatures with high accuracy and demonstrate energy savings and better occupant comfort compared to the default system controller.
\end{abstract}


\section{Introduction}

In 2018, the residential and commercial buildings accounted for about 40\% of the total U.S. energy consumption \cite{EIA2018}.
Even 1\% of the energy savings amount to \(\sim\)400 trillion Btu.
This is equivalent to reducing average power generation by 13 GW for the entire year.
Assuming we first shut down coal-fired power plants, 100 million tons less CO\(_2\) will be pumped into the atmosphere in a year.
With ever-growing energy demands, efficient energy systems, in particular with advanced control systems, can potentially make a massive positive impact on the environment.

\customspace

\noindent \textbf{Current challenges.} Control systems in buildings are mostly rule-based, and thus, they are energy and cost-inefficient.
The use of advanced control systems that replace these rules with model-based predictive control (MPC) can potentially save more than 10\% of overall energy usage \cite{Sturzenegger2016,Ma2012}.
MPC optimizes the performance of building energy systems taking into account weather forecasts and current operating conditions while maintaining occupant comfort and meeting required operation and safety constraints.
However, MPC requires a reasonably accurate model of the building, and buildings are very complex systems to model.
Thus, the traditional physics-based modeling approaches like the white-box and the grey-box techniques, although detailed, are cost and time prohibitive \cite{Sturzenegger2016}. 
As the building characteristics change with time, the model identification must also be repeated to update the model.
Moreover, such an expensive and complex modeling procedure is unique to every building. 
For all these reasons, physics-based modeling of large scale buildings suffers from practical challenges.

\customspace

\noindent \textbf{Combining machine learning with controls.}
A promising direction that addresses the aforementioned challenges in the modeling of buildings focuses on using black-box models for predictive control.
The data required to learn these models are obtained from building automation systems (BAS) that store historical logs from sensors and multimeters.
In the past, different machine learning algorithms have been used to learn black-box models.
\cite{JainCDC2017,JainTCPS2018,SmarraAE2018,SmarraADHS2018} focus on regression tree and random forest based approaches for MPC with experiments in EnergyPlus.
\cite{Buenning2019} tested this random forest based approach on a real building.
\cite{JainICCPS2018} focus on the use of Gaussian processes for experiment design, stochastic MPC, and online model updates with EnergyPlus as a test-bed.
\cite{Zhang2019} applied model-based reinforcement learning on a building model in EnergyPlus.
\cite{Gao2016} used sensitivity analysis (and not MPC) with deep neural networks for data center cooling.
While data-driven MPC (MPC based on black-box models in place of physics-based) reduces the engineering effort and time required to build white and grey-box models, it poses several other challenges such as (1) quality of historical data required to learn black-box models, and (2) computational complexity of the optimization problem if the learned models are non-convex or non-differentiable, see \cite{JainICCPS2018}.

\customspace

\noindent \textbf{Related work.}
In the literature, neural networks have been used for either modeling or MPC or both in different ways.
A small survey that classifies these approaches can be found in \cite{Afram2017}.
Since most of them focus on simulated environments, we discuss here the ones that tested control performance on a real building. 
In \cite{Afram2017}, neural networks are used to train temperature models and control effort is minimized (instead of energy consumption) for a residential HVAC system.
In \cite{Huang2015}, physics-based MPC is used to control a building in Terminal 1 of the Adelaide Airport.
A `neural network feedback linearization method' is then used to convert decision variables (energy supply) in the MPC problem to actual inputs (cooling valve operation and outdoor air damper) to the building.
In the following work \cite{Huang2015b}, neural networks are used to implement an optimal start-stop control rule to guarantee thermal comfort, but energy optimization is not considered.
To the best of the authors' knowledge, this paper shows for the first time, on a real building, that neural networks can be used to represent both energy and temperature dynamics in MPC to trade-off energy usage and occupant comfort.

\customspace

\noindent \textbf{Contributions.}
This paper makes the following contributions.
First, we present an approach for predictive control based on neural networks.
Using historical data from the building automation system and the weather station, we learn different neural networks that predict energy usage and zone temperatures, and then set up optimization for energy management that allows us to trade-off between energy usage and temperature setpoint tracking.
Second, we demonstrate the efficacy of our approach on a 2-story building in L'Aquila, Italy, that is equipped with a heating system from Mitsubishi.
We show in our experiments that through supervisory control, we can reduce the energy usage while keeping occupants comfortable without any modification to the existing heating system.
Third, we introduce the underlying tool \(\mathtt{tf-ipopt}\) that enables constrained nonlinear optimization in TensorFlow to solve the above MPC problem.
\section{From predictive modeling to predictive control}
\label{S:inversion}

Our goal is to learn neural networks of the form
\begin{align}
y_{t+1} = f\left( y_{t}, y_{t-1}, \dots, y_{t-\delta_y}, d_{t}, d_{t-1}, \dots, d_{t-\delta_d}, u_{t}, u_{t-1}, \dots, u_{t-\delta_u} \right),
\label{E:control}
\end{align}
where output \(y\) is either energy or temperature of one of the zones, \(d\) represents the disturbances, \(u\) the control variables.
The dynamic behavior of the output variables is captured by the lagged terms.
The order of auto-regression denoted by  \(\delta_{\{y,d,u\}}\) are hyperparameters chosen during cross validation.
In the learning step, we restructure the time series of \(y\), \(d\) and \(u\) obtained from raw sensor data to create data samples at each time instance in the above format.
For optimal decision making, we use model \eqref{E:control} in the MPC problem as follows
\begin{align}
& \minimize_{u_{0}, \dots, u_{N-1}} \sum_{t=0}^{N-1} \ \ (y_{t+1}\!-\!y_{\mathrm{ref}})^2 ~+~ {u_{t}}^T R {u_{t}} \label{E:mpc:generic} \\
& 
\begin{aligned}
\st \ \ 
& y_{t+1} = f \left( y_{t}, y_{t-1}, \dots, y_{t-\delta_y}, d_{t}, d_{t-1}, \dots, d_{t-\delta_d}, u_{t}, u_{t-1}, \dots, u_{t-\delta_u} \right),\\
& u_{t} \in \mathcal{U},  \nonumber \\
& \forall t \in \{0,\dots,N-1\}, \nonumber \\
\end{aligned}
\end{align}
where \(N\) is the control horizon, \(R \succ 0\) is the cost matrix, \(y_{\mathrm{ref}}\) is the reference to be tracked.
While \eqref{E:mpc:generic} is an example of a tracking controller, the cost function can be changed depending upon the application.

Using this approach, for a two-story building described in Section~\ref{S:redhouse}, we formulate the MPC problem in Section~\ref{S:mpc}, and present experimental results in Section~\ref{S:results}.
In our case, neural networks are trained in TensorFlow \cite{Tensorflow2015} with a stochastic optimization solver -- Adam \cite{Kingma2014}, and the optimization \eqref{E:mpc:generic} is solved with a non-linear interior point optimization solver -- IPOPT \cite{Waechter2009}.
\section{Building}
\label{S:redhouse}
\begin{figure}[t]
	\centering
	\includegraphics[width=0.48\columnwidth]{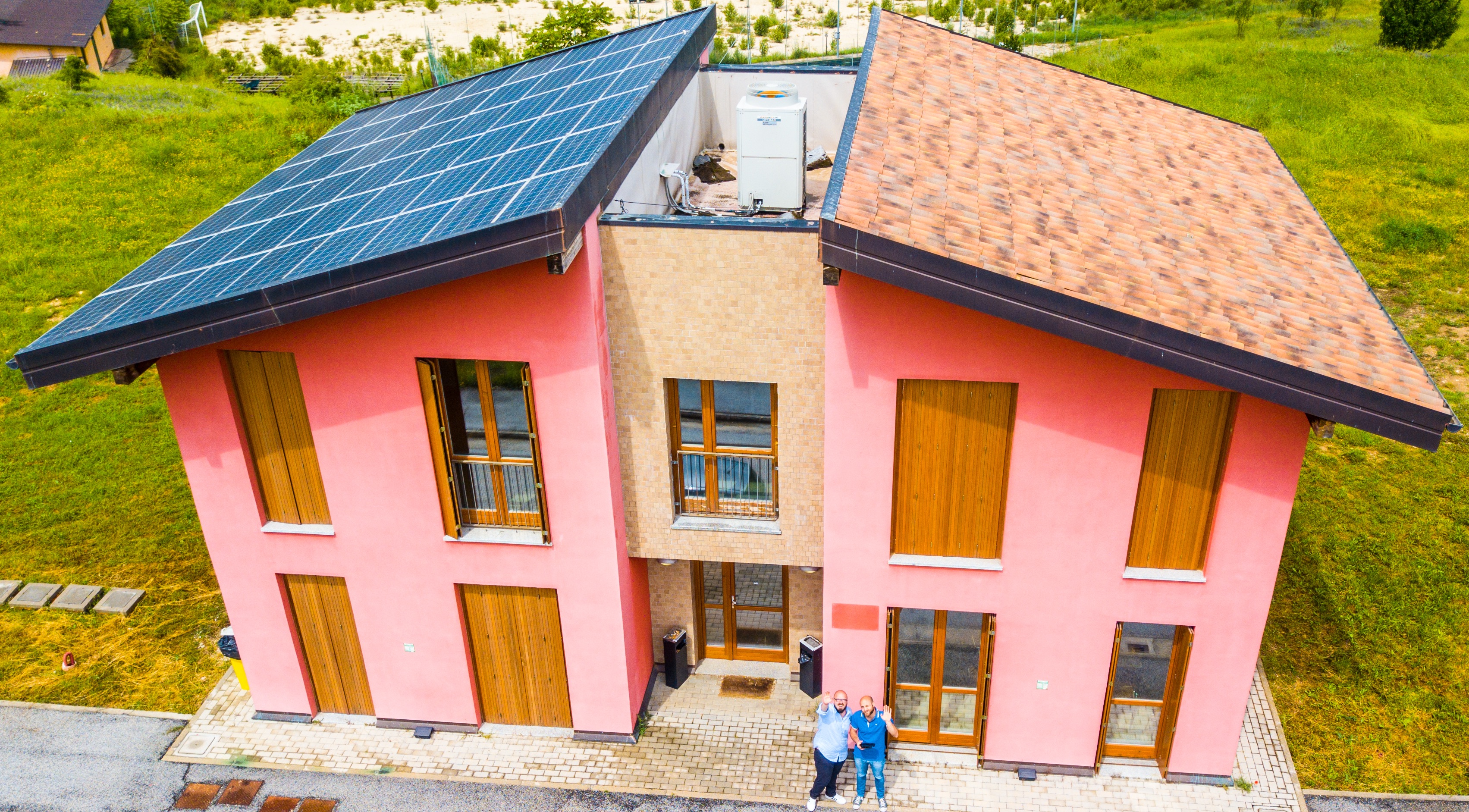}
	\hspace{10pt}
	\includegraphics[width=0.43\columnwidth]{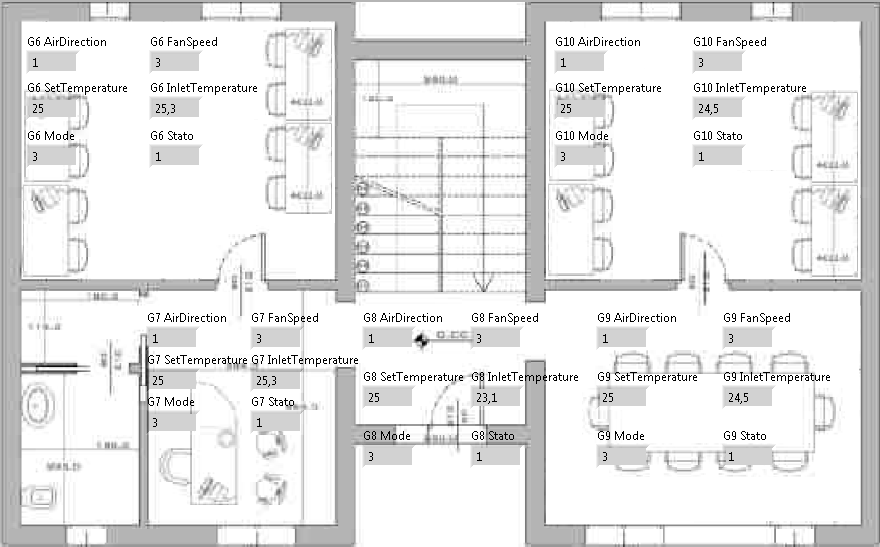}
	\caption{Left: external view of the building at University of L'Aquila, Italy. 
Right: layout of the first floor.}
	\vspace{-10pt}
	\label{F:redhouse}
\end{figure}

We consider a two-story building with a rooftop unit, as shown on the left in Figure~\ref{F:redhouse}.
The building is located inside the Coppito campus of the University of L'Aquila, Italy. 
Each floor is composed of 5 zones -- 4 rooms and a small lobby, and can be independently controlled.
The layout of the first floor is shown on the right in Figure \ref{F:redhouse}.
Enlarged layouts of both floors are attached in Appendix~\ref{A:layouts}.
The gross area of the ground and first floor is 72$\text{ m}^\text{2}$ and 77$\text{ m}^\text{2}$, respectively.
\subsection{Heating system}
The building is equipped with a variable refrigerant flow heat pump (a type of rooftop unit) from Mitsubishi.
The heating system comprises of (1) an outdoor unit on the roof that includes a compressor and an evaporator, and (2) an indoor unit (also called split) in each zone that includes a fan and a condenser.
Heating is provided through refrigerant conduits connecting indoor and outdoor equipment.
The thermal energy from the evaporation and compression phases is carried by the refrigerant.
This energy is transferred by the condenser into the zones where warm air is then distributed by the fan.
Additionally, each room and lobby is equipped with a temperature sensor.
The power consumption of the building is measured using a multimeter.
The system is configured to be programmatically controlled via M-NET (Mitsubishi network) protocol.
We discuss how we extract the data from the BAS and control this building remotely in Appendix~\ref{A:dataacquisition}.

\subsection{Role of supervisory control} 
Traditional control systems in buildings rely on fixed rules for the manipulation of temperature setpoints.
For example, during Winter, the setpoints may be kept constant at 25\(^\circ\)C during working hours.
For any chosen setpoint, the low-level controllers try to keep the zone temperatures close to the chosen setpoint.
In our case, this controller is a \textit{relay} controller from Mitsubishi.
During Winter, when the setpoint in a zone is kept constant, the corresponding indoor unit is switched ON when the measured temperature is \(\sim\)1.5$^\circ$C below the setpoint.
As the zone starts to heat up, the indoor unit is switched OFF when the measured temperature exceeds the setpoint by \(\sim\)0.5$^\circ$C.
Since different rooms have different temperatures, the external unit may be kept ON for usually longer periods of time.
Now, the goal of a neural network based predictive controller is to dynamically change the setpoints based on measured temperatures and external weather conditions, in order to (1) reduce the amount of time the external unit is ON, hence reduce energy consumption, or (2) provide better thermal comfort by tracking a given setpoint and reducing the variation in the measured temperature due to bang-bang behavior of the relay controller.
In Section~\ref{S:results}, we show how the dynamic changes to setpoints help achieve the aforementioned objectives.
\section{Model predictive control with neural networks}
\label{S:mpc}
This section is divided into two parts.
In Section~\ref{SS:models}, we discuss the model training process -- features (regressors) for energy and temperature models, the architecture of neural networks, and performance validation of the trained models on unseen data sets.
We formulate the nonlinear MPC problem and describe how to solve it in Section~\ref{SS:control}.

\subsection{Predictive modeling}
\label{SS:models}

We learn different models for energy and temperature predictions.
The sampling time is chosen to be \(T_s=2\) min since the compressor, and hence the energy consumption is very responsive to changes in temperature setpoints.
The same dynamical models are used for the length of the horizon to derive energy and temperature states in the future.
The models were trained using data from the months of October 2018 -- February 2019, and October -- November 2019.
In total, 18 weeks of data were used for training (the building was non-operational in the remaining weeks), of which 12 weeks of data were obtained with random excitation (kept constant for 1-2 hours) in temperature setpoints between 22-28\(^\circ\)C and the remaining data were obtained with constant setpoints.
The building was unoccupied during the entire duration of the experimentation.

\customspace

\noindent \textbf{1.~Energy prediction.}
This model predicts the energy consumption \(E_t\) over the next sampling time, i.e., between \(\left[t,t+1\right]\) and is given by the expression
\begin{align}
E_t = f_E\left( E_{t-1}, E_{t-2}, \dots, E_{t-\delta_E}, d_{t}, d_{t-1}, \dots, d_{t-\delta_d}, u_{t}, u_{t-1}, \dots, u_{t-\delta_u} \right).
\label{E:energy}
\end{align}
We model \(f_E\) as a neural network with 2 hidden layers (50 neurons each) and Rectified Linear Unit (ReLU) activation function.
The disturbances \(d \in \mathbb{R}^{13}\) include outside temperature, humidity and solar radiation, temperature measurements from all 10 zones.
The weather data are obtained from the weather station provided by the \cite{Cetemps}.
Note that the temperature predictions for the future time steps are obtained using models \(f_{T^j}\) in \eqref{E:temp}.
The control inputs \(u \in \mathbb{R}^{11} \) include temperature setpoints for all 10 zones and the compressor mode (boolean), in that order.
We use \(\delta_E=4\), \(\delta_d=3\), and \(\delta_u=3\).
Although the compressor mode is not a free control variable since the compressor state ON/OFF is decided based on an embedded control law in the BAS microcontroller, adding it as a feature in the model drastically improves the modeling accuracy.

\customspace

\noindent \textbf{2.~Temperature prediction.} 
A separate model is learned for each room to predict the temperature in that room at time \(t+1\), given the temperature measurements from the sensor until time \(t\):
\begin{align}
T_{t+1}^j = f_{T^j}\left( T_{t}^j, T_{t-1}^j, \dots, T_{t-\delta_T}^j, d_{t}, d_{t-1}, \dots, d_{t-\delta_d}, u_{t}^j, u_{t-1}^j, \dots, u_{t-\delta_u}^j \right). \label{E:temp}
\end{align}
Here, each \(f_{T^j}\) is a neural network with only 1 hidden layer (50 neurons) and ReLU activation function \(\forall j \in \{1,2,\dots,10\}\).
The temperature dynamics is essentially given by a piecewise affine function of all the features and is sufficient to predict room temperatures with high accuracy.
The disturbances \(d \in \mathbb{R}^3\) include outside temperature, humidity, and solar radiation, and control input \(u \in \mathbb{R} \) includes the temperature setpoint for that room.
We use \(\delta_T=3\), \(\delta_d=3\) and \(\delta_u=3\).

\customspace

\begin{figure}[t]
	\centering
	\vspace{-10pt}
	\includegraphics[width=0.49\columnwidth]{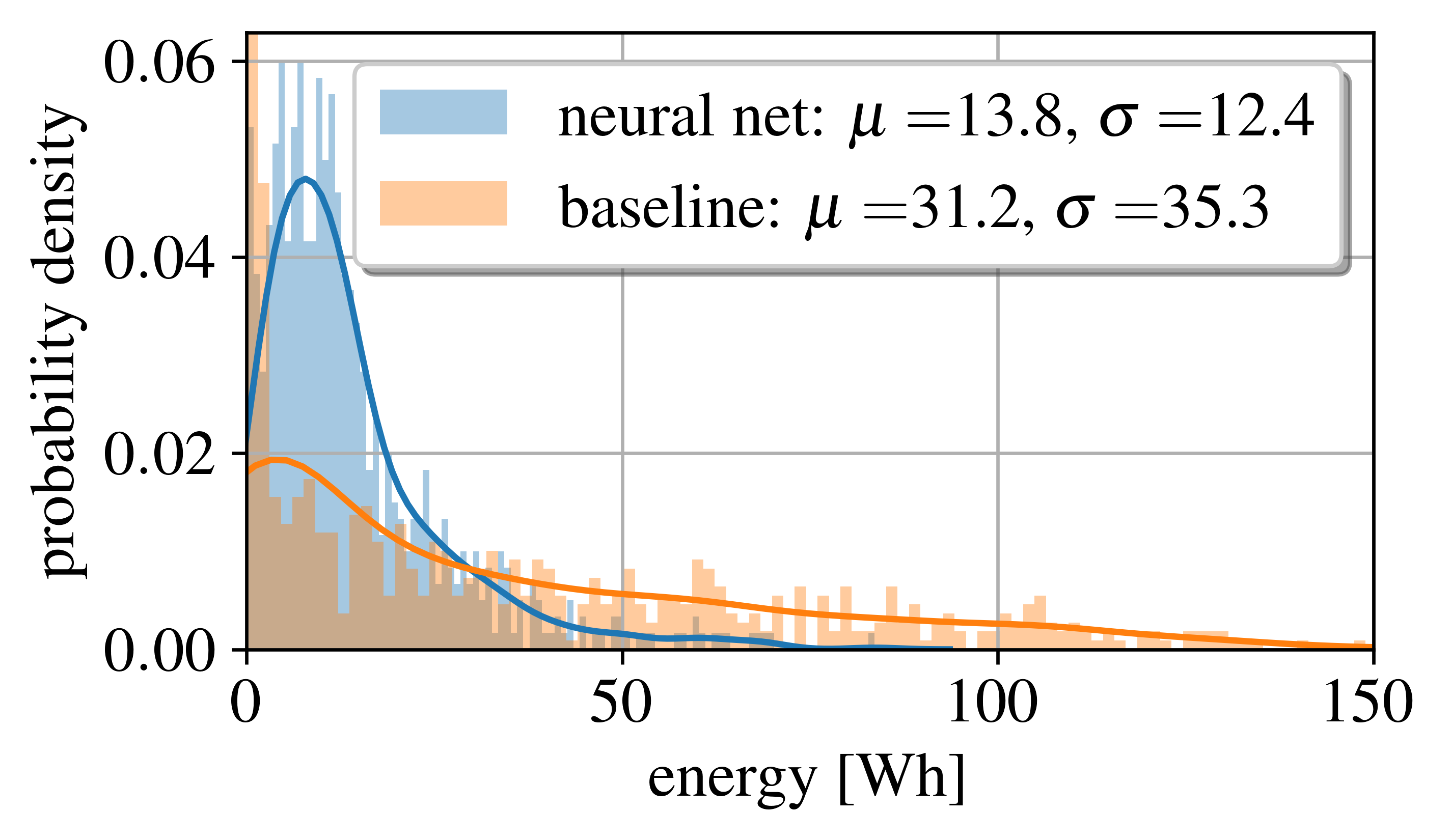}
	\includegraphics[width=0.49\columnwidth]{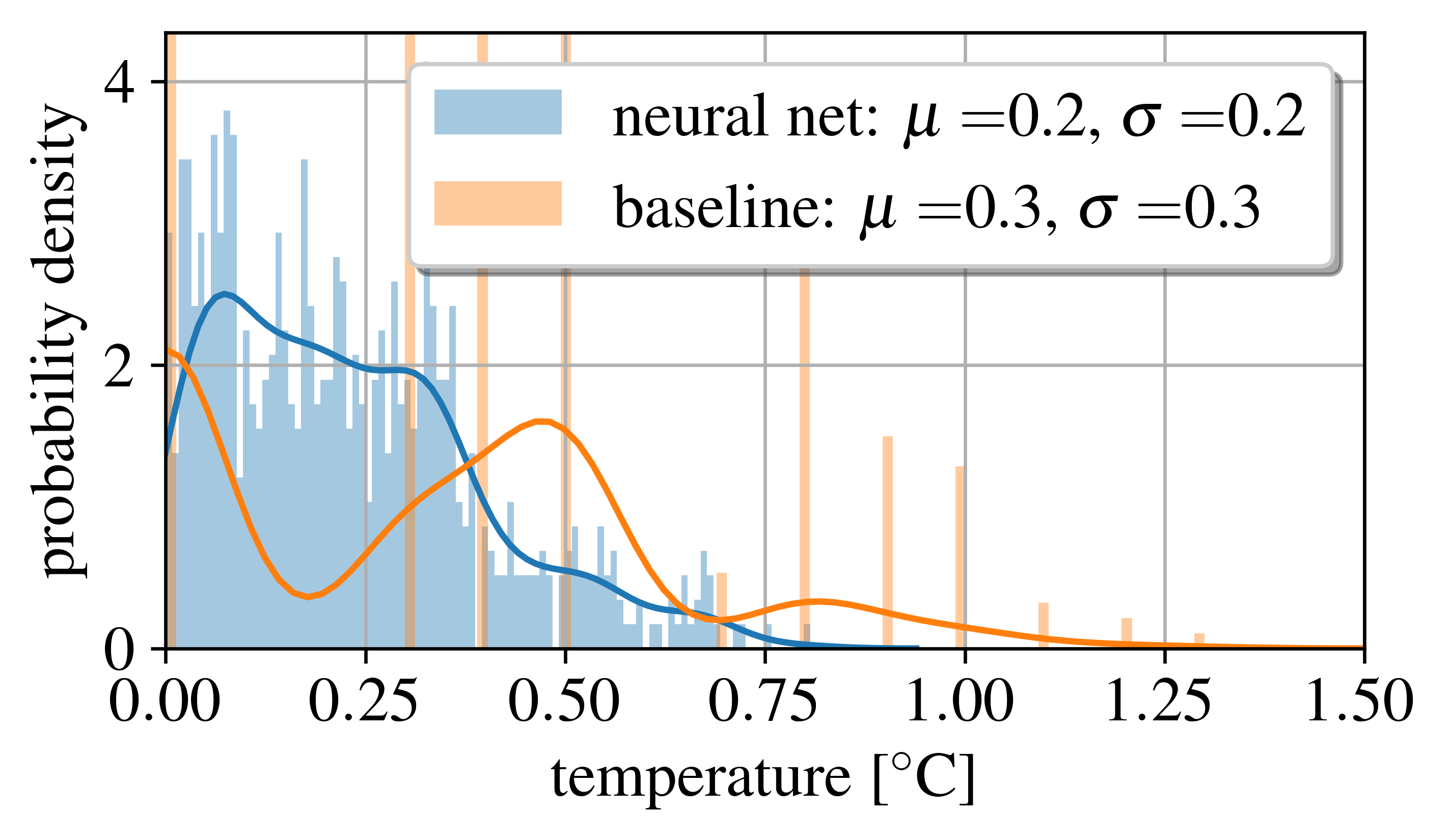}
	\vspace{-10pt}
	\caption{Absolute predictions errors for energy and temperature in one of the zones for the validation period. The curves represent approximate probability densities derived from empirical distributions.}
	\vspace{-10pt}
	\label{F:errorstats}
\end{figure}

\noindent \textbf{Model validation.} The statistics for absolute predictions errors with energy and temperature models are shown in Figure~\ref{F:errorstats}.
We compare 1-step predictions from the neural networks against a naive baseline model that assumes the predictions at the next time step are same as the measurements at the current time step, i.e.~\(E_t = E_{t-1}\) in \eqref{E:energy} and \(T_{t+1}^j = T_{t}^j\) in \eqref{E:temp}.
We observe that the baseline prediction errors show heavy tails even with a small sampling time of 2 min while the probability densities for the neural networks are concentrated in the region with small errors.
Thus the predictions from the neural network are more robust.
This is expected, especially for energy consumption due to fast dynamics of the compressor.
Note that we observe discrete behavior in the temperature plot for the baseline case because the measurements are available with only one decimal precision.

\subsection{Receding horizon control}
\label{SS:control}

The control problem is set up as a finite receding horizon optimization.
The dynamical models derived in Section~\ref{SS:models} serve as equality constraints over the horizon.
The BAS accepts only integral values for temperature setpoints.
Since solving a mixed-integer non-linear program is much harder and computationally challenging to solve, we solve an approximate problem with continuous input space and then round off the solution of the optimization to the nearest integers.
More precisely, at each time step, we solve an optimization problem that allows us to trade-off energy usage and setpoint tracking
\begin{align}
& \minimize_{u_{0}, \dots, u_{N-1}} \ \ \sum_{t=0}^{N-1} \left( \lambda_E~{E_{t}} ~+~ \sum_{j=1}^{10} \ \lambda_T~(T_{t+1}^j\!-\!T_{\mathrm{ref}}^j)^2 ~+~ 100~\epsilon_t^j \right) \label{E:mpc} \\
& 
\begin{aligned}
\st\ \ 
& E_t = f_E\left( E_{t-1}, E_{t-2}, \dots, T_{t-\delta_E}, d_{t}, d_{t-1}, \dots, d_{t-\delta_d}, u_{t}, u_{t-1}, \dots, u_{t-\delta_u} \right), \nonumber \\
& T_{t+1}^j = f_{T^j}\left( T_{t}^j, T_{t-1}^j, \dots, T_{t-\delta_T}^j, d_{t}, d_{t-1}, \dots, d_{t-\delta_d}, u_{t}^j, u_{t-1}^j, \dots, u_{t-\delta_u}^j \right), \nonumber \\
& u_{\mathrm{min}} \leq u_{t} \leq u_{\mathrm{max}},  \nonumber \\
& T_{\mathrm{min}}^j - \epsilon_t^j, \leq T_{t+1}^j \leq T_{\mathrm{max}}^j + \epsilon_t^j, \  \epsilon_t^j \geq 0, \nonumber \\
& \forall t \in \{0,1,\dots,N-1\}, \  \forall j \in \{1,2,\dots,10\}.  \nonumber
\end{aligned}
\end{align}
Here \(u_{t}^j\) is the \(j^{th}\) element of \(u_{t}\) \(\forall j \in \{1,2,\dots,10\}\) and slack variables \(\epsilon\) are added to prevent infeasibilities.
The sampling time for the models is 2 min but the MPC problem is solved every 5 min, using the same inputs for the next 5 min to avoid changing temperature setpoints too frequently.
The control horizon \(N\) is chosen to be 10 steps (20 min).
In Section~\ref{S:results}, we show results for 2 scenarios: (1) energy minimization only by setting \(\lambda_E=1\), \(\lambda_T=0\) and (2) temperature tracking for better occupant comfort by choosing \(\lambda_E=0\), \(\lambda_T=1\), and \(T^j_{\mathrm{ref}}=\)25\(^\circ\)C.

The optimization requires 5s to solve using \(\mathtt{tf-ipopt}\), a custom tool we built for constrained optimization in TensorFlow using IPOPT.
IPOPT is an open source software package for large-scale nonlinear optimization \cite{Waechter2009}.
To solve general non-linear (possibly non-convex) optimization problems like \eqref{E:mpc} which have inequality constraints, we needed an interface that allows us to call IPOPT from TensorFlow since the energy and temperature models were trained in TensorFlow.
The tool is available at \url{https://github.com/jainachin/tf-ipopt}.
An alternate option is to use CasADi \cite{Andersson2018}.
Note that training of machine learning models involving minimization of a loss function are unconstrained optimization problems.
\section{Experiments}
\label{S:results}

We evaluate MPC problem \eqref{E:mpc} under two different scenarios and compare the results against a baseline controller that chooses fixed setpoints. 
The following three controllers are compared.

\customspace

\noindent \textbf{1.~Baseline: relay controller with fixed setpoints.}
This is a relay controller that comes with the heating unit.
Constant setpoint of 25\(^\circ\)C is chosen for each zone.

\customspace

\noindent \textbf{2.~MPC-min: energy minimization only.}
Set \(\lambda_E=1\) and \(\lambda_T=0\) in \eqref{E:mpc}. 
The goal is to minimize energy consumption while keeping all zone temperatures between 23\(^\circ\)-27\(^\circ\)C.

\customspace

\noindent \textbf{3.~MPC-tracking: setpoint tracking for better occupant comfort.}
Set \(\lambda_E=0\), \(\lambda_T=1\), and \(T^j_{\mathrm{ref}}=\text{25}^\circ\)C in \eqref{E:mpc}. 
This MPC controller adjusts the setpoints to keep the measured temperature closer to the chosen reference.

\subsection{MPC-tracking versus Baseline}
\label{SS:mpctrackingvsbaseline}
Baseline controller was tested on December 13-15, 2019 (blue) and MPC-tracking on December 16-18, 2019 (green).
Recall that Baseline controller has a maximum threshold cutoff 0.5$^\circ$C above and minimum threshold 1.5$^\circ$C below the setpoint, i.e.~the relay width is \(\sim\)2$^\circ$C.
This asymmetry by design is aimed to save energy and prevent excess heating.
The comparison of occupant comfort (quality of temperature tracking) is shown in Figure~\ref{F:control}.
The mean and standard deviation of measured zone temperatures are \(\mu=\text{24.8}^\circ\text{C}\), \(\sigma=\text{0.4}^\circ\text{C}\) for MPC-tracking and \(\mu=\text{24.3}^\circ\text{C}\), \(\sigma=\text{0.8}^\circ\text{C}\) for Baseline.
In the case of Baseline controller, the mean zone temperature is far from 25$^\circ$C (high bias) and also shows large fluctuation (high variance).
By dynamically changing the setpoints, MPC-tracking keeps the zone temperature closer to 25$^\circ$C (low bias) and also reduces the fluctuation significantly (low variance).
High bias with Baseline is attributed to the asymmetry in the design of the relay controller (see above) which can be reduced by tracking 25.5$^\circ$C instead.
However, to reduce the variance, the relay width must be reduced from 2$^\circ$C to \(\sim\)1$^\circ$C.
This is only possible by changing the design of the existing heating system.
MPC-tracking provides the same benefit without any modifications.

\subsection{MPC-min versus Baseline}
\label{SS:mpcminvsbaseline}
\begin{figure}[t]
	\centering
	\vspace{-15pt}
	\includegraphics[width=\columnwidth]{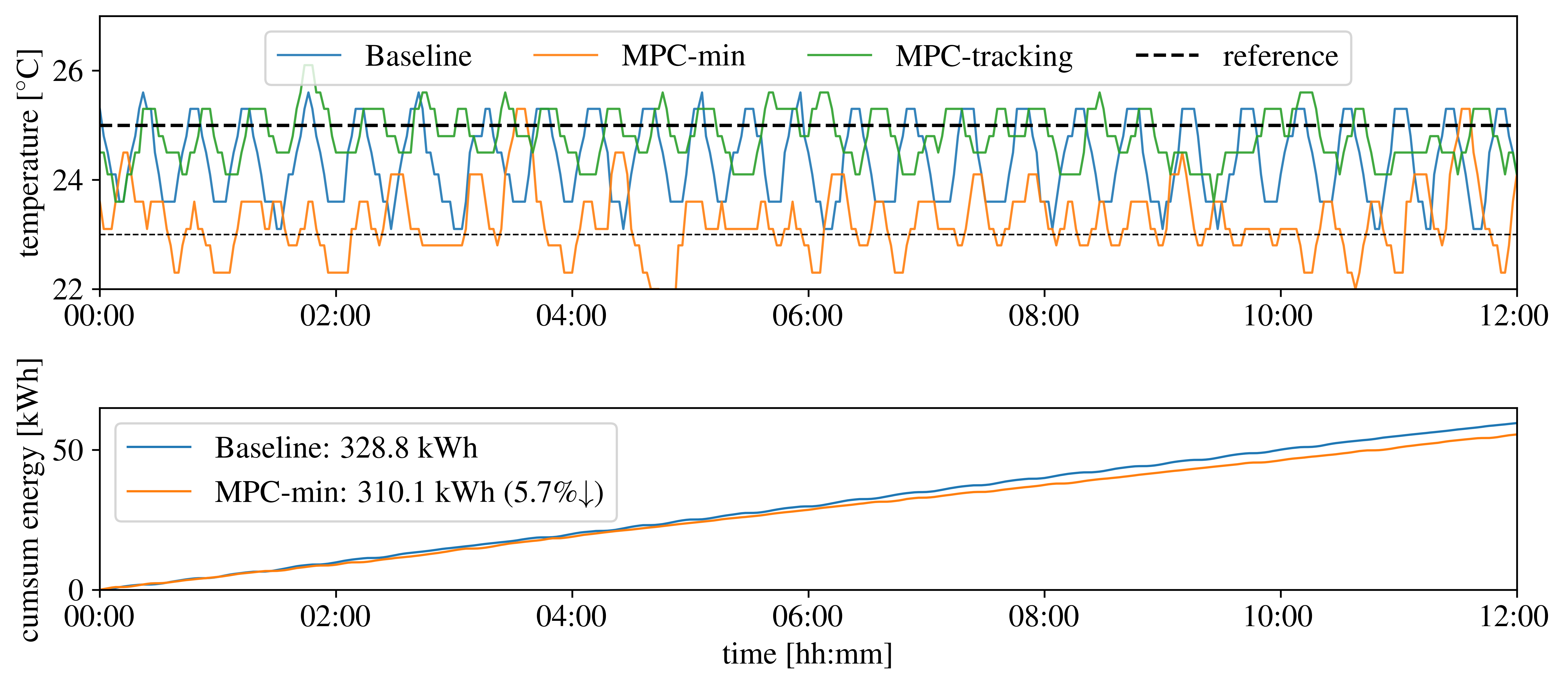}
	\vspace{-25pt}
	\caption{Comparison for temperature in one of the zones (top) and energy savings (bottom). MPC-tracking provides much better occupant comfort compared to Baseline. At the end of 3-day experiment, MPC-min consumed 5.7\% less energy compared to Baseline by keeping the zone temperature closer to lower bound.}
	\vspace{-5pt}	
	\label{F:control}
\end{figure}

\begin{figure}[t]
	\centering
	\includegraphics[width=0.32\columnwidth]{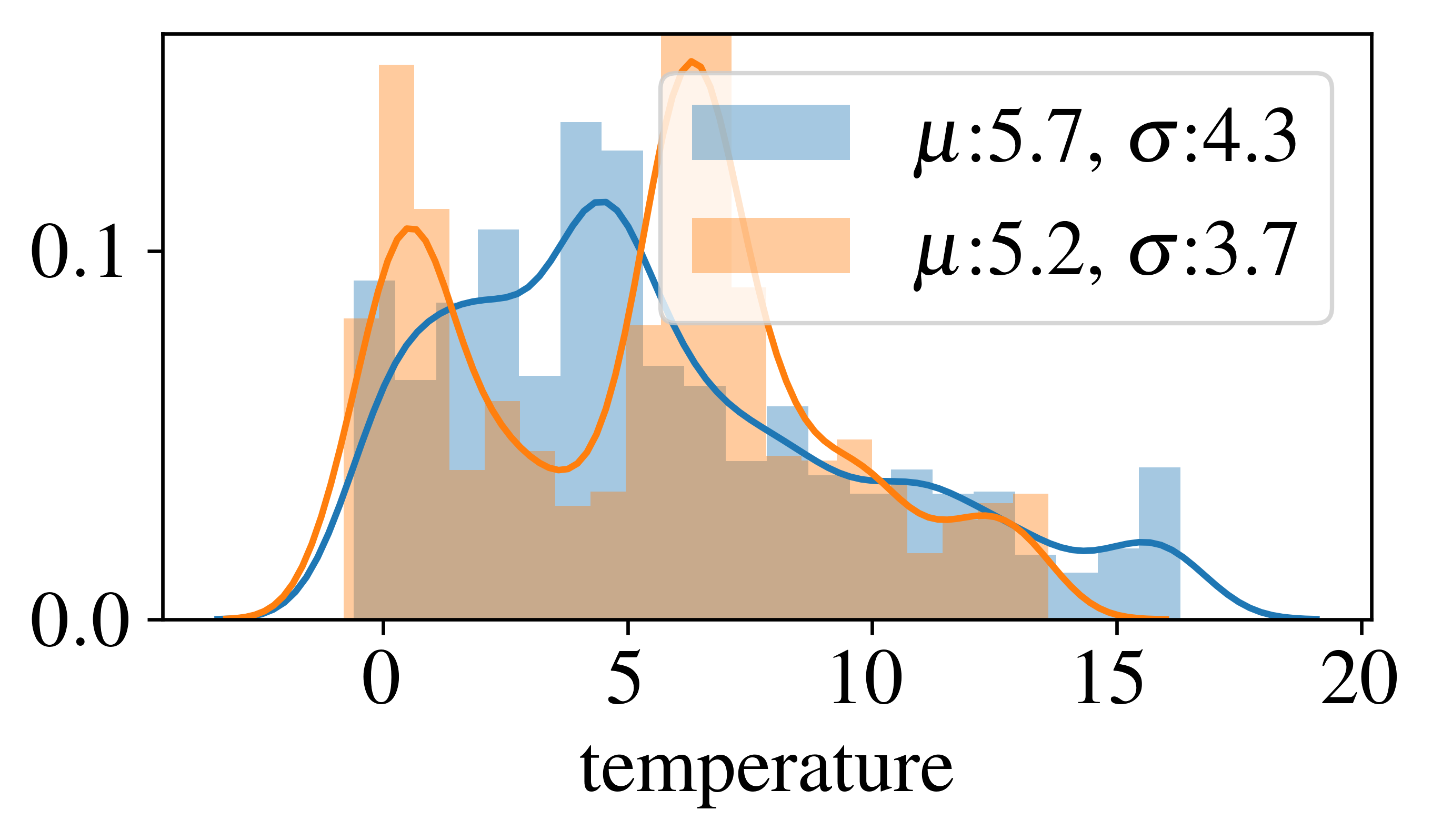}
	\includegraphics[width=0.32\columnwidth]{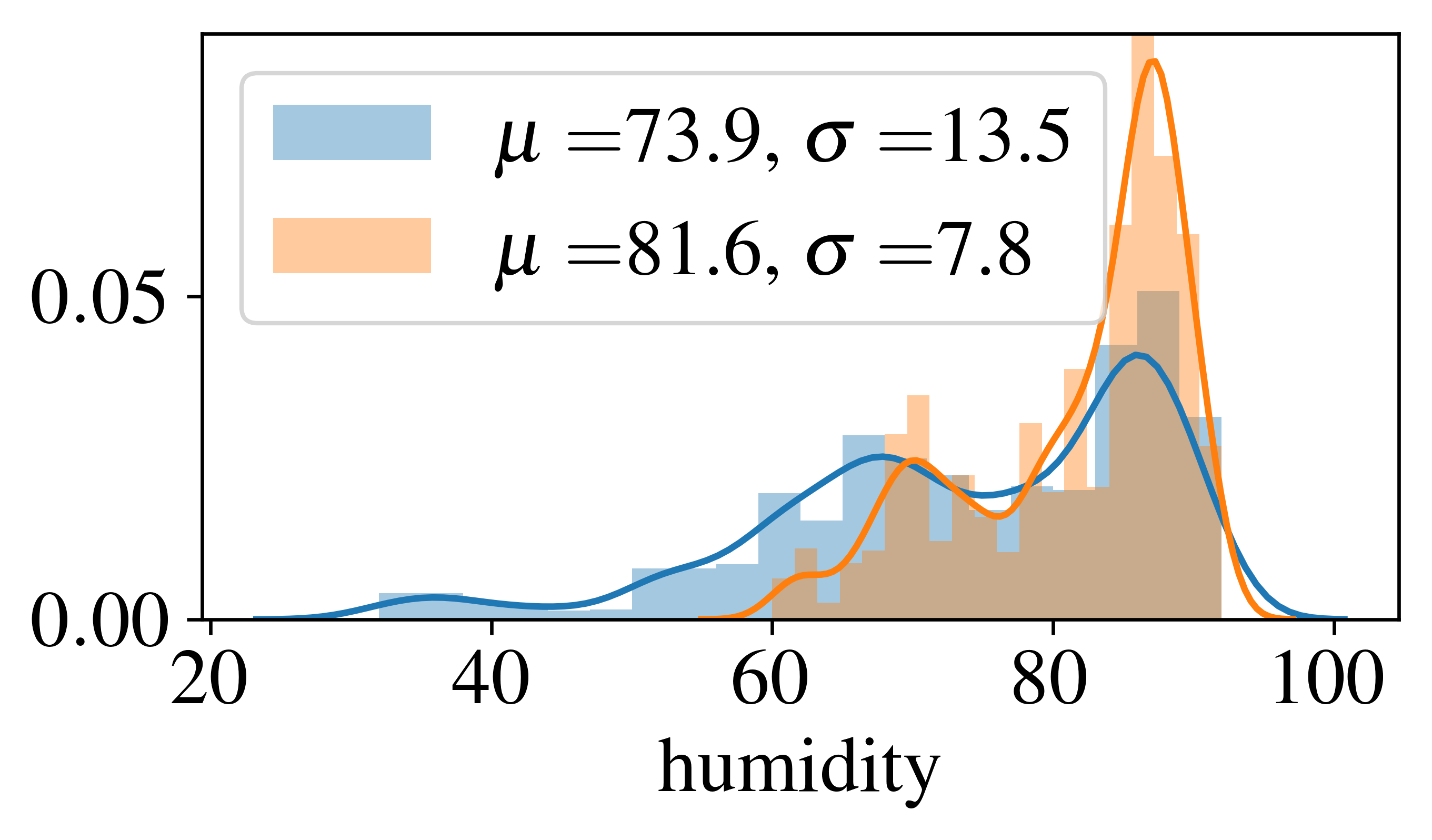}
	\includegraphics[width=0.32\columnwidth]{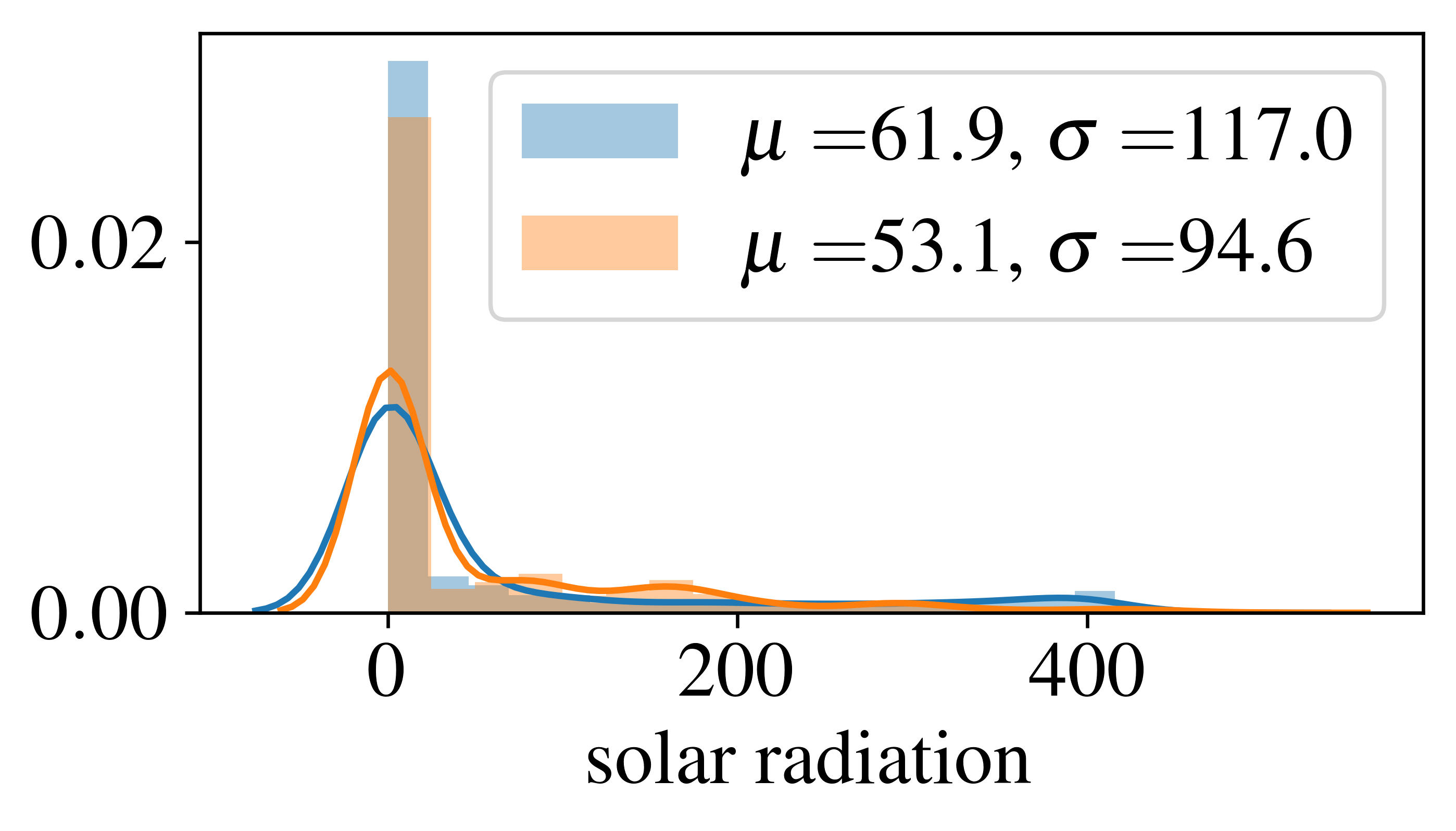}
	\vspace{-10pt}
	\caption{Weather statistics on the days of comparison. Baseline controller was tested on December 13-15, 2019 (blue) and MPC-min on December 5-7, 2019 (orange).}
	\label{F:weatherstats}
\end{figure}

To evaluate the benefits of MPC-min against Baseline from Section~\ref{SS:mpctrackingvsbaseline}, it is important to consider similar weather conditions, especially outside temperature.
However, since only one controller can be tested at a time, a perfect comparison with the exact same weather conditions is impossible.
We identified two periods of three consecutive days with similar weather conditions to compare MPC-min with Baseline.
The distributions of weather disturbances on these days are shown in Figure~\ref{F:weatherstats}.
MPC-min was run on December 5-7, 2019 (orange).
On average, it is warmer and less humid on the days Baseline controller was run.
A comparison for energy consumption is shown in Figure~\ref{F:control} (bottom).
As expected, MPC-min pushes the zone temperature closer to the lower bound at 23\(^\circ\)C to realize energy savings, see Figure~\ref{F:control} (top).
Due to the short horizon and internal system dynamics, sometimes the zone temperature goes below the lower bound.
For clarity, we show results for only the initial 12 hour period from the days of experiments.
We observe that the difference between the cumulative energy between Baseline and MPC-min is continuously increasing with time.
At the end of 3 days, MPC-min consumes 18.7 kWh less energy, a 5.7\% decrease over Baseline.
We also note that the days when Baseline was tested were warmer, so less heating will be required to achieve the same setpoint.
Further, there is a potential for Baseline controller to reduce energy consumption if we change the reference temperature below 25$^\circ$C.
However, this will come at the expense of much higher variance in the zone temperature as we discussed in Section~\ref{SS:mpctrackingvsbaseline} and thus more constraint violations.
On the other hand, MPC-min serves as a supervisory controller reducing energy usage and fluctuation in zone temperature without any modifications to the heating system.

\customspace

\noindent 
In applications like demand response and peak demand reduction where optimal decisions are sensitive to varying electricity pricing, model-based predictive control will outperform a ruled-based strategy (like fixed setpoints) by a significant margin since the setpoints will need to be dynamically changed to reduce energy costs.
Thus, our approach for neural networks based MPC would stand out against Baseline.
Two such examples problems are designing (1) MPC-min for minimizing energy \textit{costs} and (2) MPC-tracking for following a reference demand response signal.
\section{Conclusion and future work}
\vspace{-5pt}
We present an approach to learning neural networks that predict energy consumption and zone temperature dynamics in a two-story building in L'Aquila, Italy, equipped with a heating system from Mitsubishi.
We set up nonlinear MPC problem using these neural networks that allows us to trade-off energy savings and better occupant comfort.
By dynamically changing the temperature setpoints (supervisory control), the controller reduces energy consumption while respecting comfort bounds.
In a separate experiment, we also show that we can achieve better occupant comfort by reducing the variance in temperature tracking without any modifications to the existing heating system in the building.
Our results make a strong case for the application of MPC with black-box models, in particular, neural network-based models in building energy optimization.

The following are exciting directions for future research.
First, study the effect of dynamic electricity pricing and how we can exploit the building's thermal inertia to save energy costs.
Second, evaluate the performance of our controller in providing custom occupant comfort, i.e., tracking different temperatures in different rooms.
Third, investigate the continual learning of neural networks using model-based reinforcement learning (RL).
As building properties and weather conditions change with time, the goal is to minimize the maintenance of neural networks required in manual functional tests by leveraging the exploration capabilities in RL.


\newpage
\appendix
\section{Floor level layouts of the building}
\label{A:layouts}
\begin{figure}[h]
	\centering
	\includegraphics[width=0.9\columnwidth]{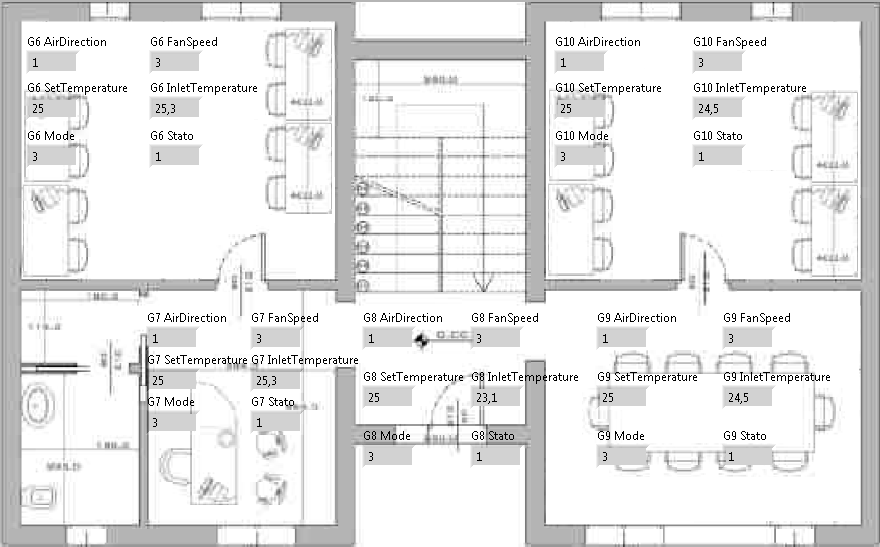}
	\caption{Layout of the ground floor.}
	\label{F:redhouseGroundFloor}	
\end{figure}

\begin{figure}[h]
	\centering
	\includegraphics[width=0.9\columnwidth]{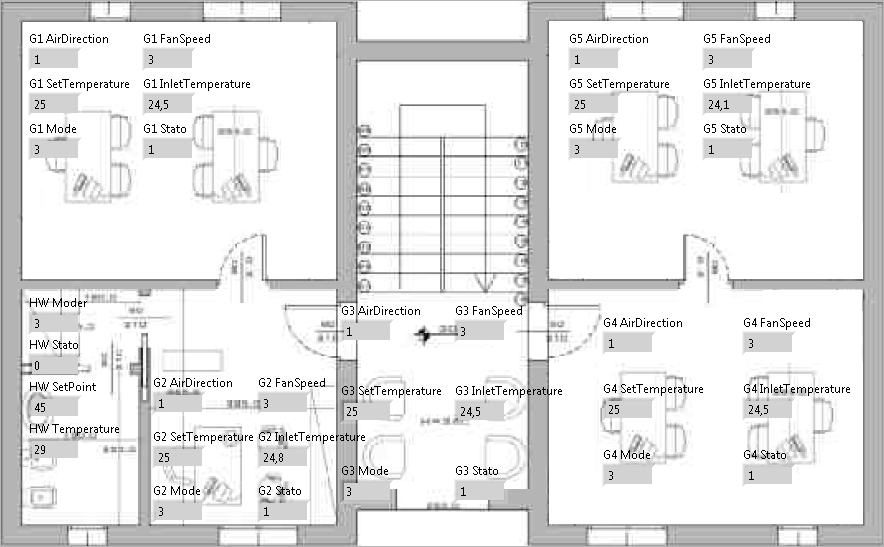}
	\caption{Layout of the first floor.}
	\label{F:redhouseFirstFloor}		
\end{figure}

The ground floor is shown in Figure~\ref{F:redhouseGroundFloor} and the first floor in Figure~\ref{F:redhouseFirstFloor}.

\section{Data acquisition}
\label{A:dataacquisition}
The building data acquisition system consists of 3 major components:

\customspace

\begin{figure}[h]
	\centering
	\includegraphics[width=\columnwidth]{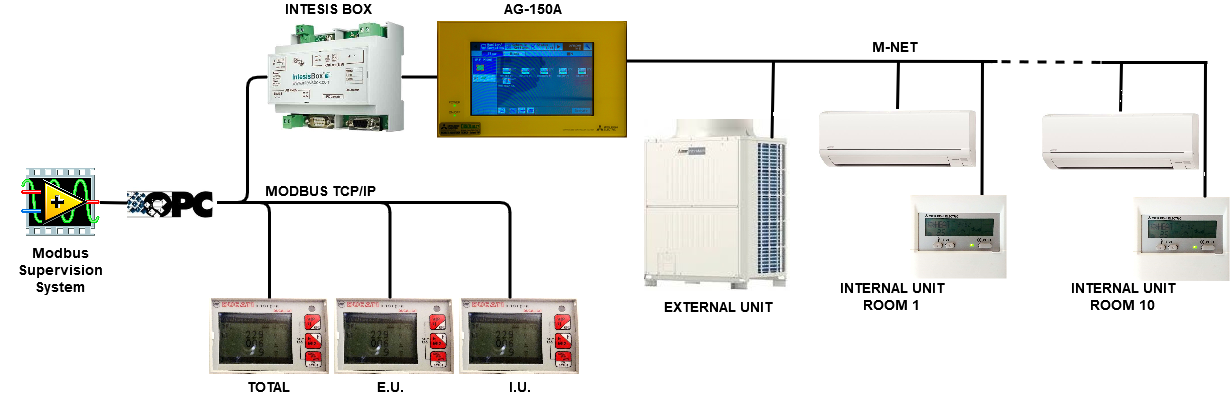}
	\caption{Links between Mitsubishi and Ducati devices with LabVIEW via Modbus TCP/IP network.} 
	\label{F:MNETtoMODBUS}
\end{figure}

\noindent \textbf{1.~Local server} -- This hosts a LabVIEW application that reads and writes data to Mitsubishi's building automation system via Modbus TCP/IP protocol.
The building's HVAC system is based on Mitsubishi proprietary serial network called M-NET. 
In order to communicate with the sensors via LabVIEW, it was necessary to add a translator from M-NET to Modbus TCP/IP to allow OPC to act as a link between LabView and the system tags.
The complete scheme is shown in Figure~\ref{F:MNETtoMODBUS}.

\customspace

\noindent \textbf{2.~Remote Elasticsearch database in AWS cloud} -- This stores real-time logs for remote monitoring and visualization using Grafana. 
We use \cite{Elasticsearch2019}, a distributed, RESTful search and analytics engine capable of handling terabytes of data together with \cite{Grafana2018}, an open sourced tool for analytics and real-time monitoring.
This particularly helps in running control experiments and visualizing data in real-time without being physically present in the building.
The experiments for functional testing and MPC send setpoints for each room to ``controller'' index in Elasticsearch.
Once the database is updated, the most recent setpoints are sent to the local server by the communication link below.

\customspace

\noindent \textbf{3.~Link between local server and remote database} -- The purpose of this link is to sync data between the LabVIEW application and the Elastisearch database every 15 seconds.
When the setpoints are updated in the ``controller'' index in Elasticsearch, it fetches the data and sends them to the LabVIEW application that further relays them to the BAS via Modbus TCP/IP protocol.
Simultaneously, the measurements from the sensors and multimeters are read from the BAS and sent to ``measurements'' index in Elastisearch database.


\newpage
\bibliography{ms}

\end{document}